\documentclass[12pt,epsf]{article}
\usepackage{graphicx}
\usepackage{amsmath}
\usepackage{enumerate}
\usepackage{eufrak}
\def\lsim{\raise0.3ex\hbox{$\;<$\kern-0.75em\raise-1.1ex\hbox{$\sim\;$}}}
\def\gsim{\raise0.3ex\hbox{$\;>$\kern-0.75em\raise-1.1ex\hbox{$\sim\;$}}}

\newcommand{\beq}[1]{\begin{equation}\label{#1}}
\newcommand{\bea}[1]{\begin{eqnarray}\label{#1}}
\newcommand{\eea}{\end{eqnarray}}
\newcommand{\be}{\begin{equation}}
\newcommand{\ee}{\end{equation}}
\newcommand{\ba}{\begin{eqnarray}}
\newcommand{\ea}{\end{eqnarray}}
\newcommand{\barr}{\begin{array}}
\newcommand{\earr}{\end{array}}

\newcommand{\rarr}{\rightarrow}
\def\21{$SU(2) \otimes U(1) $}

\def\rf#1{(\ref{#1})}

\newcommand{\mx}{\left[\begin{array}}
\newcommand{\finmx}{\end{array}\right]} 
\newcommand{\mxp}{\left(\begin{array}} 
\newcommand{\finmxp}{\end{array}\right)} 

\def\mathbf#1{\hbox{\bf #1}}
\def\textrm#1{\hbox{#1}}
\def\half{{\textstyle{1 \over 2}}}

\newcommand {\ignore}[1]{}

\newcommand {\gpp}{g_{\phi\phi}}
\newcommand {\gxx}{g_{xx}}

\begin{document}
\vspace*{-1in}
\renewcommand{\thefootnote}{\fnsymbol{footnote}}
\begin{flushright}
\texttt{
} 
\end{flushright}
\vskip 5pt
\begin{center}
{\Large{\bf 
Closed timelike curves in asymmetrically warped brane universes
}}
\vskip 25pt 

{\sf
Heinrich P\"as$^{1}$, Sandip Pakvasa$^{2}$, James Dent$^{3}$, Thomas J. Weiler$^{3}$}
\vskip 10pt
{\it \small  
$^1$~Fakut\"at f\"ur Physik, Technische Universit\"at Dortmund, D--44221 Dortmund, Germany}\\
{\it \small  
$^2$~Department of Physics \& Astronomy, 
University of Hawaii at Manoa,
2505 Correa Road, Honolulu, HI 96822, USA}\\
{\it \small
$^3$~Department of Physics and Astronomy, 
Vanderbilt University, Nashville, TN 37235, USA}\\
\today

\vskip 20pt

{\bf Abstract}
\end{center}

\begin{quotation}
{\small
In asymmetrically warped spacetimes different warp factors are assigned to
space and to time.
We discuss causality properties of 
these warped brane universes
and argue that scenarios with two extra dimensions
may allow for timelike curves which can be
closed via paths in the extra-dimensional bulk. 
In particular, necessary and sufficient conditions on the  
metric 
for the existence of closed
timelike curves are presented.
We find a six-dimensional warped metric which satisfies the CTC conditions,
and
where
the null, weak and dominant energy conditions are satisfied on the brane
(although only the former remains satisfied in the bulk).
Such scenarios are interesting, since they open the possibility of 
experimentally testing the chronology protection conjecture
by manipulating on our brane initial conditions of gravitons or 
hypothetical gauge-singlet  fermions 
(``sterile neutrinos'') which then 
propagate
in the extra dimensions.
}
\end{quotation}

\vskip 20pt  

\setcounter{footnote}{0}
\renewcommand{\thefootnote}{\arabic{footnote}}


\section{Introduction}
\label{sec:intro}
The physics of time travel has fascinated science fiction aficionados
and scientists alike. In particular, the seminal papers 
of Morris, Thorne and Yurtsever \cite{morris} on traversable wormholes 
initiated 
a considerable research library of serious attempts to 
transmit information to the past,
i.e. to generate closed timelike curves (CTCs).
Several spacetime settings, mostly contrived or 
oversimplified in some way, have been
discussed in the literature.  These include G\"odel's rotating universe 
\cite{goedel}, the rotating cylinder of
van Stockum and Tipler \cite{stockum,tipler}, 
Gott's pair of moving 
cosmic strings \cite{gott}
\footnote{Compare, however, the arguments given in \cite{deser}.}, 
Wheeler's spacetime foam \cite{foam},
regions inside the horizon of
Kerr- and Kerr-Newman geometries \cite{kerr}, Alcubierre's 
warp drive \cite{warp}, and Ori's vacuum torus \cite{ori}. 
Typically these spacetimes suffer from obstacles
of either unphysically fast rotation to tip the Lorentz cones,
or the requirement of exotic matter with negative energy density
which violates the so-named null, weak, strong and 
dominant energy conditions.  
Several analyses indicate possible instabilities of such spacetimes to
classical perturbations and/or quantum fluctuations \cite{instab}.
This situation has inspired Stephen Hawking's
``chronology protection conjecture'' \cite{hawking}, which states that 
the ultimate laws of physics prevent the appearance of CTCs. 
Hawking's ``chronology horizon'' \cite{hawking} is 
a special type of Cauchy horizon, which separates spacetime regions 
where CTCs occur from spacetime regions where chronology is protected.
Although apparently plausible situations seem naively to violate causality,
quantum corrections to the stress-energy tensor diverge 
in semi-classical calculations
at the chronology horizon.
It has been argued that the back-reaction to the metric would 
destroy the potential time-machine on the horizon.

Whether Hawking's chronology protection conjecture holds 
beyond the semi-classical
treatment, so that chronology is truly protected, is still not known.
Very probably a better understanding of quantum gravity will be necessary to 
resolve this issue in the future.
In the meantime, a study of physics under the unusual conditions 
surrounding the chronology horizon 
may provide more insight into chronology protection.
One might glimpse some fascinating new physics 
proposed to avoid the obvious paradoxes associated with time travel.
These paradoxes include the Grandfather and Bootstrap paradoxes.
In the Grandfather paradox, one modifies the initial 
conditions that lead to one's own existence;
in the Bootstrap paradox, an effect is its own cause. 
If the chronology protection conjecture is false, even more wonderful 
discoveries 
may await the serious researcher.
Proposals 
include non-Hausdorff manifold geometry
\cite{visser}, 
where the same event has 
multiple futures or pasts,
and the many-world interpretation of quantum mechanics, 
with switching between parallel histories~\cite{deutsch}.

The advent of theories with large extra dimensions 
has provided yet new room for chronology violations (see
e.g. the discussion in \cite{strings}).
Extra dimensions were originally motivated by the consistency of string theory.

More recently, large (or even infinite)
extra dimensions have been discussed as a possible new way
to understand the hierarchy problem ($M_{\rm weak}\ll M_P$) 
\cite{xtra,RS} (for reviews see e.g. \cite{xtrarevs})
and to keep neutrino masses small \cite{dvali}. 
In many extra-dimension models, ordinary Standard Model (SM) 
fields are confined on a brane (our three-surface), while
gravitons and other hypothetical SM singlet fields
are allowed to propagate also in the 
extra-dimensional bulk.
A generic feature of such spacetimes seems to be the existence of
signals, mediated by the graviton or SM singlets, 
taking ``shortcuts'' through the extra dimension.
As viewed from our brane world, these shortcuts appear as 
superluminal communication \cite{kaelb}.
Such apparent superluminal communication, via graviton shortcuts in the 
bulk \cite{freese},
or earlier, via wormholes \cite{hoch}, 
has been proposed as a possible
solution to the cosmological horizon problem 
(relaxing one of the needs for an inflationary epoch in the early universe). 
While there seems to be agreement in the literature that extra
dimensional spacetimes admit bulk shortcuts under rather generic conditions, 
whether 
these shortcuts solve the horizon problem
depends on the details of the specific extra-dimensional 
model~\cite{ishihara,caldwell,stoica,abdalla}. 

In this paper we discuss
causality violations arising in a particularly interesting class of 
extra-dimensional scenarios allowing for bulk shortcuts, so-called
asymmetrically warped spacetimes. 
Asymmetric warping assigns a different warp factor to time versus space 
coordinates.
We work with lightlike world-lines
accessing extra-dimensions, 
but we expect our conclusions to apply to any extremely relativistic quanta 
which have access to the extra-dimensional bulk
(thus using the term closed timelike curve (CTC) as being interchangeable
to closed lightlike curve).
Examples 
of such quanta are the graviton and the 
hypothetical
gauge-singlet ``sterile'' neutrino.

We begin the CTC story by reviewing 
closed timelike curves in a prominent class of spacetimes including 
the G\"odel- and Tipler-van-Stockum (GTvS) spacetimes in Section~2.
In Section~3 we extend the discussion of possible CTCs to higher-dimensional 
spacetimes,
and derive the general conditions on the metric which allow CTCs.
We discuss the causality properties of asymmetrically warped 5-dimensional 
spacetimes in Section~4, 
and show that CTCs do not occur (unless spacetime has the topology of a
flat torus).
In contrast, we show in Section~5 that closed timelike curves can exist 
within a 
6-dimensional generalization of asymmetrically warped spacetimes.  
We explicitly find a metric with  
two warped extra space dimensions which allows CTCs.
In Section~6 we discuss the energy conditions for the 6D metric with 
CTCs.
A brief discussion of the role that sterile (gauge-singlet) neutrinos 
may play in the communication with the past or future 
and a recapitulation and discussion complete this paper 
in Section~7.

\section{CTCs in G\"odel and  Tipler-van-Stockum spacetimes}
\label{sec:GTvSspacetimes}
Before discussing the causality properties of asymmetrically warped
spacetimes,
it is instructive to review
two prominent early examples of spacetimes implementing closed timelike
curves. 
The G\"odel metric describes a pressure-free perfect fluid with
negative cosmological constant and rotating matter, and the 
Tipler-van-Stockum~(TvS)
spacetime is being generated by a rapidly rotating infinite cylinder.
In both cases the metric can be written as

\be
\label{stockum}
ds^2= + g_{tt}(r)\,dt^2 +2 g_{t\phi}(r)\,dt d\phi - g_{\phi\phi}(r)\,d\phi^2 
  - g_{rr}\,dr^2 - g_{zz}\,dz^2\,.
\ee
Here the $g_{\mu \nu}$ are complicated functions of 
the radial distance $r$ from the symmetry axis, and a parameter 
characterizing the 
angular velocity of the cylinder (not shown).
The azimuthal coordinate $\phi$ assumes values on the interval 
$\phi \in [0,2\pi\}$.
Writing $g_{tt}\,dt^2+2g_{t\phi}\,dtd\phi-g_{\phi\phi}\,d\phi^2$ as 
$(g_{tt}+\Omega^2\,g_{\phi\phi})\,dt^2-g_{\phi\phi}\,(d\phi-\Omega\,dt)^2$,
with $\Omega=g_{t\phi}(r)/g_{\phi\phi}(r)$, 
makes it clear that the G\"odel and TvS metrics have an interpretation 
in terms of a rotating coordinate system with a radially-dependent angular 
speed $\Omega(r)$.
The sign of the rotation is positive if $g_{t\phi}$ and $g_{\phi\phi}$
have the same relative sign, and negative if the relative sign is opposite.

To guarantee a local Minkowskian metric at every spacetime point, 
on and off the brane, 
we maintain the Lorentzian signature.
The 4D metric has Lorentzian signature provided that 
$g_4\equiv Det(g_{\mu\nu})<0$.
For $g_{rr}\,g_{zz}>0$ as in the G\"odel and TvS (GTvS) metrics,
this condition becomes
\be
\label{signature}
(g_{tt}\,g _{\phi\phi} + g_{t\phi}^2) > 0.
\ee 
A dynamical approach to GTvS causality examines the 
purely azimuthal null-curve with $ds^2=0$.
One gets 
\be{}
\dot{\phi}_\pm= \frac{g_{t\phi} 
   \pm \sqrt{g^2_{t\phi} + g_{tt}\,g_{\phi\phi}}}{g_{\phi\phi}}\,,
\ee 
where the $\pm$ refers to co-rotating and counter-rotating lightlike signals.
The coordinate time for a co-rotating path is 
\be{}
\Delta T_+ = \Delta\phi 
\left(
\frac{g_{\phi\phi}}
   {g_{t\phi}+\sqrt{ g_{t\phi}^2 + g_{\phi\phi}\,g_{tt}}}
\right)\,.
\ee
As $g_{\phi\phi}$ goes from positive to negative, the light-cone tips 
such that the
azimuthal closed path is traversed in negative time
\be{}
\Delta T_+ = \frac{-2\pi\,|g_{\phi\phi}|}{g_{t\phi} +\sqrt{ g_{t\phi}^2 
+g_{\phi\phi}\,g_{tt}}}
\quad\quad [g_{\phi\phi} <0] \,.
\ee
The quantum returns to its origin before it left,
marking the existence of a CTC.

Note that the Lorentzian signature is maintained even as $g_{\phi\phi}$ 
switches sign 
as long as the argument of the square-root, proportional to $-g_4$, 
remains positive definite. 
Note also that the Lorentzian signature is maintained as $g_{\phi\phi}$ 
switches to a negative value 
as long as $g_{t\phi}> \sqrt{-g_{\phi\phi}\,g_{tt}}$.  
In particular, $g_{t\phi}$ cannot be zero.
In the following we will apply a similar argument to different scenarios
of asymmetrically warped spacetimes.

For completeness, we give the analogous result for the 
counter-rotating light signal.  
The period of counter-rotation is 
\be{}
\Delta T_- = \frac{2\pi\,(g_{t\phi}+\sqrt{-g_4}\,)}{g_{tt}}\,.
\ee

If $g_{tt}$ switches its sign at some $r_0$, then $\Delta T_-$ becomes 
negative. 
However, such a sign change either produces singular behavior in 
$\Delta T_-$ or requires a 
discontinuous behavior in $g_{tt}$.
In the GTvS models, $g_{tt}$ does not change sign.

The clear discriminator of the arrows of time are the 
slopes of the local light-cone,
\beq{s+-}
s_{\pm}(r)=(r\dot{\phi}_\pm)^{-1} = \frac{1}{r}\  
\frac{g_{\phi\phi}}{g_{t\phi}\pm\sqrt{-g_4}}
   =-\frac{1}{r}\  \frac{g_{t\phi}\mp\sqrt{-g}}{g_{tt}}\,.
\ee
Notice that if $g_{\phi\phi}$ and $g_{tt}$ are positive, 
then regardless of the sign of $g_{t\phi}$,
the light-cones (worldlines) remain in the first and 
second quadrants of the $(t,\phi)$ plane
(as is the case of the Minkowski light-cone).
Thus, for a backward flow of time, $g_{\phi\phi}$ (or, $g_{tt}$) must 
go through zero and become negative.

It is useful to consider the product of slopes
\beq{ss}
s_+(r)\,s_-(r) = \frac{-1}{r^2} \frac{g_{\phi\phi}}{g_{tt}}.
\ee
For time to move backwards one of the world lines defining the light-cone 
must move into the lower half of the $t-\phi$ plane.
From (\ref{ss}) one can see that 
(i) this happens smoothly if $g_{\phi\phi}$ 
goes through zero; (ii) happpens discontinuously if $g_{tt}$ goes through zero;
(iii) that a smooth change in the sign of $g_{t\phi}$ cannot move either slope 
through zero to the  domain of negative time.

With the focus here on a smooth change of sign for $g_{\phi\phi}$, 
it is useful to examine the slopes at small $\gpp$.
One finds 
\beq{s+-small}
s_\pm ({\rm leading\ order\ in\ }\gpp) = 
\left\{
\barr{r}
\frac{1}{2r}\ \frac{\gpp}{g_{t\phi} }\\ 
\\
-\frac{2}{r}\ \frac{g_{t\phi}}{g_{tt}}
\earr
\right.
\ee
It is clear that the slope $s_+$ goes through zero with $\gpp$, 
leaving the first quadrant and moving into the fourth quadrant.
With increasing $\phi$, time for the associated co-rotating world 
line runs backwards.
On the other hand, the sign of $s_-$ remains unchanged, and time for the 
associated counter-rotating 
world line continues to run forward. 
In the following we will apply similar arguments to different scenarios
of asymmetrically warped spacetimes.

It is instructive to mention the visceral arguments against the 
relevance of the G\"odel and TvS metrics.
First of all, they are not asymptotically flat, and so presumably cannot 
occur within our Universe;
rather, they must be our Universe, which contradicts observation.
Secondly, the initial conditions from which they can evolve are either 
non-existent~(G\"odel) 
or sick~(TvS).
Furthermore, the TvS metric assumes an infinitely-long cylinder of matter,
which is unphysical.
On the positive side, literally, the Einstein equation endows 
$\rho=T^0_{\ 0}=(R^0_{\ 0}-\half R)/8\pi\,G_N$ (with the geometric RHS 
determined by the metric) 
with a positive value everywhere;
there is no need for ``exotic'' $\rho<0$ matter.
A further positive feature is the simplicity of finding the CTC 
by travel along the periodic variable $\phi$. 
We will later revisit these pluses and minuses with our 
results for asymmetrically warped spacetimes.

\section{Extending GTvS
spacetimes to extra dimensions}
\label{sec:extendingST}
In this section we will model 6D spacetimes after the GTvS spacetimes.  
We label the two additional space dimensions $u$ and $v$,
and we ``warp'' the bulk by letting the metric coefficients of the familiar
spacetime coordinates $(t,{\vec x})$ depend on the bulk coordinates $u$ 
and $v$.

\subsection{A periodic path off the brane}
\label{sub:periodicpath}
By construction, the causal properties of the metric depend 
only on the bulk coordinates $u$ and $v$.
Therefore, we are led to consider first a path off the brane 
whose projection onto the brane is periodic.
Such a path mimics closely that of GTvS\@.
We begin with the 6D line element and metric
\beq{extendedST2}
ds^2=g_{tt}(u,v)\,dt^2+2g_{t\phi}(u,v)\,dt\,(rd\phi)
-g_{\phi\phi}(u,v)\,(r\,d\phi)^2-dr^2-du^2-dv^2\,;
\ee
the brane coordinate $\theta$ is not needed, and so we have 
set it equal to $\pi/2$.
For this line element we have explicitly displayed the powers of $r$ so that
all elements of $g_{\mu\nu}$ are dimensionless.
The Lorentzian signature is maintained everywhere as long as 
$-g_6\equiv -Det(g_{\mu\nu}) = (g_{tt}\,g_{\phi\phi}+g_{t\phi}^2)\,r^2 > 0$ 
everywhere.

The algebra yielding the travel time for a periodic path in the hyperslice 
$(u,v)=(u_1,v_1)$ 
is little changed from the GTvS prototype.
For fixed $r\equiv r_1$ we readily arrive at the travel time along a 
co-rotating path:
\beq{periodicDeltaT}
\Delta T_+=r_1\Delta\phi\,\left(
\frac{g_{\phi\phi}(u_1,v_1)}{g_{t\phi}(u_1,v_1)+\sqrt{-g_6(u_1,v_1)}}
\right)\,.
\ee
We take $\Delta\phi>0$ and $g_{t\phi} >0$.
The transit time for the periodic co-rotating path 
is therefore negative if $g_{\phi\phi}(u_1,v_1)$ is negative.
Since the periodic path may be transited an arbitrary number of times, 
the finite time 
required for the lightlike quanta to travel from the brane at $(u,v)=(0,0)$ 
to the hyperslice in the bulk at $(u,v)=(u_1,v_1)$ and back can be neglected.
Taking $\Delta\phi$ to be a multiple of $2\pi$, one obtains a CTC.

We have little intuition for a globally-defined, differentially-rotating 
coordinate system.
On the other hand, it may be possible to construct a coordinate system with 
rotation over a finite volume,
e.g., by embedding a Kerr-like solution in the 6D space.
However, such a construction, if possible at all, is a complication beyond 
the scope of this paper.
We choose to leave the concept of a relative rotation between brane and 
bulk to future study, and instead consider next  
a related but different metric.

\subsection{A linear path off the brane}
\label{sub:linearpath}

We may replace the periodic coordinate of GTvS with the unbounded $x$ 
coordinate,
and omit the $y$ and $z$ coordinates for brevity.
Then one obtains 
\beq{extendedST1}
ds^2=g_{tt}(u,v)\,dt^2+2g_{tx}(u,v)\,dxdt-g_{xx}(u,v)\,dx^2-du^2-dv^2\,.
\ee
(Notice in particular the sign convention on the coefficient of $dx^2$.)

The speed of light at any point will depend on 
$(u,v)$ through the metric elements.
The restriction to Lorentzian signature implies that
\beq{6Dsignature}
-g_6 \equiv -{\rm Det}(g_{\mu\nu}) = 
g_{tt}(u,v)\,g_{xx}(u,v)+g_{tx}^2(u,v)>0\,.
\ee
World lines for lightlike travel (null lines) satisfy
\beq{GTvSnull}
0=g_{tt}(u,v)+2g_{tx}(u,v)\,{\dot x}-g_{xx}(u,v)\,{\dot x}^2
-{\dot u}^2-{\dot v}^2\,.
\ee
The solutions to (\ref{GTvSnull}) for the analogs of 
co-rotating and counter-rotating 
light speed at fixed $(u,v)$ are
\beq{velocities1}
{\dot x}_\pm =\frac{g_{tx}(u,v) \pm \sqrt{-g_6}}{g_{xx}(u,v)}\,.
\ee
On the brane, ${\dot x}$ must equal $c=1$, so we again choose 
$g_{tt}(0,0)=g_{xx}(0,0)=1$ and $g_{tx}(0,0)=0$.

Let us examine more closely the causal implications of Eq.~\rf{velocities1}.
We assume that $g_{tt}$ is everywhere positive, so that 
(i) coordinate time $t$ is everywhere timelike,
and (ii) no singularities are introduced in $s_+ s_-$ or in $g_{tt}$.
As shown in Section~(\ref{sec:GTvSspacetimes}), the sign of 
$g_{tx}$ does not influence the causal structure, and 
for definiteness we take it  to be positive semidefinite.
It is the sign of the metric element $g_{xx}$ that has smooth causal 
significance.

Similar to the causal analysis of the GTvS model of 
Section~(\ref{sec:GTvSspacetimes}),
we  write the two slopes of the light-cone as
\beq{s+-us}
s_{\pm} (u,v) = (\dot{x}_\pm )^{-1} = \frac{\gxx (u,v)}{g_{tx}(u,v) 
\pm\sqrt{g_{tx}^2(u,v) + \gxx (u,v) g_{tt}(u,v)}}\,.
\ee
From this, one readily gets
for the slopes (given in Eq.~(\ref{s+-us}))
\beq{ss-us}
s_+ s_- = \frac{-\gxx}{g_{tt}}\,.
\ee
It is easily seen that when $\gxx$, $g_{tt}$, and $g_{tx}$ are all positive, 
the slopes are of opposite sign, and are connected to the Minkowski metric 
in the smooth limit $g_{tx}\rarr0$.
Thus, with  $g_{tt}$ and $g_{tx}$ assumed positive, time flows in the 
usual manner if $\gxx$ is positive.
Furthermore, with $g_{xx}>0$, we have ${\rm sign}(g_{tx}\pm\sqrt{-g_6})= \pm$,
so that from Eq.~(\ref{velocities1}), 
one has ${\dot x}_+ >0$, and ${\dot x}_- <0$.
Thus, a positive $g_{xx}$ (as in the Lorentz metric) 
offers the standard situation 
with time flowing forward and velocity ${\dot x}$ having either sign. 

On the other hand, if $g_{xx}$ is negative, then 
Eq.~(\ref{ss-us}) shows that one light-cone slope has changed sign.
The small $\gxx$ limit of the slopes 
\beq{s+-small-us}
s_\pm ({\rm leading\ order\ in\ }\gxx) = 
\left\{
\barr{r}
 \frac{\gxx}{2g_{tx} }\\ 
\\
- \frac{2 g_{tx}}{g_{tt}}
\earr
\right.
\ee
reveals that it is the positive slope which has passed through 
zero to become negative,
signifying a world line moving from the first quadrant, through the $x$-axis, 
into the fourth quadrant where times flows backwards for increasing $x$.
With both slopes negative, one has that ${\dot x}_\pm (g_{xx}<0)<0$.
Thus, travel with increasing time is in the negative~$x$ direction,
while travel with decreasing time is in the positive~$x$ direction.
We summarize the causal properties of the metric~\rf{extendedST1} 
in Table~\ref{table:velocities}.


\begin{table}
\centering
\begin{tabular}{|c||c|c|} \hline
                &  $g_{xx}>0$     &  $ g_{xx}<0$ \\ \hline\hline
  $\Delta T >0$ &  ${\dot x}_+ >0 $ & \\
  	                  & ($\Delta x >0$)    & \\ 
	\cline{2-3}
                & ${\dot x}_- <0$   &  ${\dot x}_- <0$ \\ 
                & ($\Delta x <0$) &  ($\Delta x <0$)  \\ \hline
  $\Delta T <0$ & & $\dot{x}_+ <0$ \\
 &  &  ($\Delta x >0$)   \\ \hline 
\end{tabular}
\caption{ 
\label{table:velocities}
Solution types for metric~(\ref{extendedST1}), and their casual properties.
In particular, no solution exists for motion backwards in time 
along the negative-$x$ direction.}
\end{table}
 
The world line which we investigate is the following:
 the signal travels first from the brane at 
$(u,v)=(0,0)$ to the hyperslice at $(u_1,v_1)$,
then from $(u_1,v_1)$ to the hyperslice at $(u_2,v_2)$,
and finally back from $(u_2,v_2)$ to the point of origin $(0,0)$ on the brane
(see Fig.~\ref{bulkpath}).
While on the $(u_1,v_1)$ hyperslice, the signal travels a distance $\Delta X$ 
in the positive $x$-direction over a negative time $\Delta T_1= -|\Delta T_1|$.
While on the $(u_2,v_2)$ hyperslice, the signal travels back 
an equal negative distance $-\Delta X$ in time $\Delta T_2$
to close the spatial projection of the worldline on the brane.
To close the worldline on the brane, it is necessary that $T_2+T_1<0$.  
(But not equal to zero, as we allow for small positive travel times 
from the brane at 
$(u,v)=(0,0)$ to $(u_1,v_1)$, from $(u_1,v_1)$ to $(u_2,v_2)$,
and back from $(u_2,v_2)$ to $(0,0)$.)

The transit time $(\Delta T_1)_\pm$ for 
light to travel a positive distance $\Delta X >0$ 
at constant $(u_1,v_1)$, as viewed from the brane, is 
\bea{DeltaT1}
(\Delta T_1)_\pm &=& \int_0^{\Delta T_1} dt 
    = \int_0^{\Delta X} dx\ \frac{g_{xx}(u_1,v_1)}{g_{tx}(u_1,v_1)
\pm\sqrt{-g_6(u_1,v_1)}}\\ \nonumber
   &=& \Delta X\,\left(\frac{g_{xx}(u_1,v_1)}{g_{tx}(u_1,v_1)
\pm\sqrt{-g_6(u_1,v_1)}}\right)\,.
\eea
The integrations on $dt$ and $dx$ are trivial because the metric 
does not depend on the 
coordinate time $t$ or brane variable $x$.  
According to Eq.~(\ref{6Dsignature}), the Lorentz signature is 
maintained as long as 
$g^2_{tx} > g_{tt}\,(-g_{xx})$.
We have shown that the world line for $x_+$ lies below the 
$x$-axis when $\gxx <0$,
and so we require $\gxx (u_1,v_1)<0$ in order to gain negative time 
$\Delta (T_1)_+$ 
during travel on the $(u_1,v_1)$~hyperslice.
From here on, we will simply use the label $\Delta T_1$ 
for this negative $\Delta (T_1)_+$ solution 
on the $(u_1,v_1)$~hyperslice:
\beq{}
\Delta T_1\equiv \Delta (T_1)_+ =
  \Delta X\,\left(\frac{g_{xx}(u_1,v_1)}{g_{tx}(u_1,v_1)
+\sqrt{-g_6(u_1,v_1)}}\right)\,.
\ee

To close the worldline, the lightlike signal must return from 
positive $\Delta X$ 
to the origin $x=0$ in a time $\Delta T_2$ less than or equal to 
$|\Delta T_1|$.  
If this were to occur in a negative time, then we would have 
$g_{xx}<0$ and 
${\dot x}>0$.  Table~\ref{table:velocities} shows that there is 
no solution of this type 
available.  So the return path must take place in positive time, 
with ${\dot x}<0$.  
Reference again to Table~\ref{table:velocities} reveals that the 
return solution is ${\dot x}_-$.
In principle, the ${\dot x}_-$ solution on the $(u_1,v_1)$~hyperslice 
provides a return path.
However, it is easy to show that the return time $\Delta T_2$ for this 
solution 
exceeds $|\Delta T_1|$ and so fails to close the world line.
Thus, we must go to a second hyperslice at $(u_2,v_2)$.
We have 
\beq{DeltaT2} 
\Delta T_2 = \left[ \int_{\Delta X}^0 dx = -\Delta X\,\right]\, 
\left(\frac{g_{xx}(u_2,v_2)}{g_{tx}(u_2,v_2)-\sqrt{-g_6(u_2,v_2)}}\right)\,,
\ee
with $g_{xx}(u_2,v_2)$ of either sign.

The necessary condition relating the outgoing and return paths of a CTC is
that the sum $\Delta T_2 +\Delta T_1$ be less than zero.
Equivalently, the CTC conditions are that

\begin{subequations}
\label{CTCconditions}
\be
\label{CTCcondition1}
\hspace{1.0cm}
\frac{-g_{xx}(u_2,v_2)}{g_{tx}(u_2,v_2)-\sqrt{-g_6(u_2,v_2)}}
  + \frac{g_{xx}(u_1,v_1)}{g_{tx}(u_1,v_1)+\sqrt{-g_6(u_1,v_1)}}
 < 0\,, 
\ee 
and that 
\be
\label{CTCcondition2}
\hspace{3.0cm}
 g_{xx}(u_1,v_1)<0\,, 
\ee
\end{subequations}
(recall our sign convention~(\ref{extendedST1}) for $g_{xx}$).
Here $\Delta T_1<0$, $\Delta T_2>0$ has been used.
Note that (\ref{CTCcondition1}) is both necessary and sufficient, while
(\ref{CTCcondition2}) is implied by (\ref{CTCcondition1}), assuming
that the negative time $\Delta T_1$ is accumulated during the travel on the 
$(u_1,v_1)$ hyperslice. Thus (\ref{CTCcondition2})
is a necessary but not a sufficient condition. 
It is nevertheless a useful guide for our analysis
of candidates for CTC spacetimes.
The two transit times $\Delta T_1$ and $\Delta T_2$ can be made arbitrarily 
long,
and so the short-time paths from the brane to the $(u_1,v_1)$~hyperslice,
from $(u_1,v_1)$ to the $(u_2,v_2)$~hyperslice, 
and from $(u_2,v_2)$ back to the brane, can be neglected;
if we can show the existence of metric elements on the 
$(u_2,v_2)$ and $(u_1,v_1)$ hyperslices  
satisfying the constraints of Eqs.~(\ref{CTCconditions}),
we will 
have demonstrated the existence of a closed worldline for lightlike quanta.
Since $\Delta T_1$ and $\Delta T_2$ can be made arbitrarily long,
finite mass effects of order $1/\gamma^2$ may be neglected, and so 
the same conditions enable CTCs for extremely relativistic timelike quanta.

It is worth remarking that besides the necessity of the inequality 
$g_{xx}(u_1,v_1)<0$ to 
generate a negative time path, it is also necessary that 
$g_{tx}(u_1,v_1)\ne 0$.
Without this latter condition, the Lorentz signature could not be 
maintained when $g_{xx}<0$, and indeed, the square root in the second term
in~\rf{CTCcondition1} would become imaginary.

Noting that $\Delta T_1$ is negative and $\Delta T_2$ positive,
we have $\Delta T_1/\Delta X=-1/|{\dot x}_1|$ 
and $\Delta T_2/\Delta X=+1/|{\dot x}_2|$. 
Thus, we may also interpret Eqs.~\rf{CTCconditions} to say that 
\beq{velocities}
\frac{1}{|{\dot x}_2|}+\frac{-1}{{|\dot x}_1|} 
< 0\,,\quad{\rm i.e.,\ }|{\dot x}_2| > |{\dot x}_1|\,.
\ee
In words, the quantum must return to the brane at a greater speed 
than that with which it exited,
in order for the projection of the worldline onto the time axis to close.
Put another way, the slope of the world line below the $x$-axis, 
$s_1=\frac{-1}{|\dot{x}_1|}$,
must exceed in magnitude the slope of the return path, 
$s_2=\frac{1}{ |\dot{x}_2| }$.

At this point we can see that the return path, like the outgoing path, must 
occur off the brane.
The outgoing path in Eq.~(\ref{DeltaT1}) generates a negative time as 
perceived on the brane. 
This event is spacelike, superluminal, 
with the event $|\Delta T_1|/\Delta X$ outside the light-cone and therefore 
less than 1.
The inverse velocity $\Delta T_1/\Delta X = -|\Delta T_1|/\Delta X$ is 
therefore greater than $-1$.
The return path, if on the brane,
would be equal to $1/c=1$, and the inequality in (\ref{CTCcondition1}) 
would be unfulfilled.
Equivalently, we may note that a return path on the brane occurs at speed 
${\dot x}_2=1$,
whereas the superluminal outgoing path occurs at speed $|{\dot x}_1|>1$;
Eq.~\rf{velocities} is thus unfulfilled.

Since Eqs.~\rf{CTCconditions} are 
the necessary and sufficient condition for a CTC,
any metric failing to satisfy the inequalities in~\rf{CTCconditions} has no 
CTC.
On the other hand,
we have seen that the GTvS model contains a CTC in the 2+1 
dimensional
space $(r,\phi;t)$.
Thus, we expect that CTCs will populate some metrics in $N+1$ spaces, for any
$N>2$, as well. Indeed, Eq.~(\ref{CTCconditions}) summarizes the
straightforward recipe
for constructing metrics with CTCs in spaces equal to or larger than 2+1. One
simply needs
(i) a $g_{xx}$ that passes through zero as a function of another spatial
coordinate ``$u$'',
(ii) a nonzero $g_{tx}$ in the region of $u$ where $g_{xx}<0$,
and (iii) a return path suitably arranged with nonzero values $g_{xx}$ and
$g_{tx}$ in
another coordinate region of $u$. The mathematical construction of such metrics
is not in question.
What may be debated is the physics motivation for such metrics. In the 
following
sections we will develop a metric with CTCs,
motivated by a popular concept in particle physics and gravitation,
extra-dimensional ``warped'' spaces.

\section{Causality with one warped extra dimension}
\label{sec:5D}
We consider the five-dimensional
asymmetrically-warped line element with 
a single extra dimension which we label as ``$u$'':
\be{}
\label{metric}
ds^2 = dt^2 - \sum_i \alpha^2(u)\,(dx^i)^2 - du^2,
\ee 
$i=1,2,3$, with our brane located at the $u=0$ submanifold.
With no loss of generality, we may take $\alpha(u)$ to be positive.

Variants of this warped spacetime (\ref{metric}) 
can be generated by AdS-Schwarzschild or
AdS-Reissner-Nordstr\"om black holes in the bulk
\cite{chung,csaki}, and have been proposed
as solutions to the cosmological horizon problem \cite{freese},
and as a possible way around Weinberg's no-go
theorem for the adjustment of the cosmological constant \cite{csaki}. 
They also have been discussed in the context of 
the gravitational generation of cosmic acceleration \cite{pad1}, and
infrared modification of gravity \cite{pad2}.
Very recently it has been shown that sterile neutrinos propagating
in such a spacetime can account for the LSND neutrino oscillation evidence,
without the problems faced by 
conventional four-dimensional four-neutrino 
scenarios~\cite{paes}.

The warped spacetime of~(\ref{metric}) allows shortcut geodesics 
connecting spacelike-separated events on the brane if 
$|\alpha(u)|< |\alpha(0)|$ for any $u\ne 0$.
However, the metric (\ref{metric}) exhibits a global time function $t$.
Thus, taken by itself this spacetime 
is causally stable and does not allow for CTCs. 

The failure of~(\ref{metric}) to support a CTC can also be seen 
in our CTC equations~\rf{CTCconditions}.
Since $g_{xx}=\alpha^2$ in \rf{metric} cannot be negative 
without violating the assumed Lorentzian signature,
the CTC condition \rf{CTCcondition2} cannot be satisfied.

Given that the metric \rf{metric} does allow 
spacelike geodesics (as viewed from the brane),
a boosted observer may see a negative time for the outgoing path.
It is of pedagogical value to investigate \rf{metric} in the coordinates
of this boosted observer.
This effort will serve as a precursor for a successful construction 
of a metric with CTCs in six dimensions in the next section.

The metric in (\ref{metric}) is in 
Gaussian normal form with respect to $u$ 
(i.e., $g_{tu}=g_{x_i u}=0$),
so the induced metric on each hypersurface 
with constant $u$ is simply given by the
extra-dimensional metric evaluated on the hypersurface.
These induced metrics are purely Minkowskian, 
albeit with a different constant limiting velocity 
$c(u)=\alpha^{-1}(u)$ on each hypersurface.
This means that a Lorentz symmetry can be defined for each hypersurface,
but each hypersurface's Lorentz symmetry will not hold on any other 
hypersurface,
as we now discuss.

It is natural to choose $c(u=0)=1$ 
such that the induced metric on the brane is given by 
$ds^2_{\rm brane}=dt^2-dx^2$.
There then follows the usual Lorentz symmetry under the familiar 
transformations on our brane:
\be{}\label{ltrans}
x'=\gamma\left(x-\beta t\right),
~~~~t'=\gamma\left(t-\beta x\right),
~~~~u'=u=0,
\ee
or equivalently, the inverse transformation
\be{}\label{ltransinv}
x=\gamma\left(x'+\beta t'\right),
~~~~t=\gamma\left(t'+\beta x'\right),
\ee
with the usual definition $\gamma=(1-\beta^2)^{-1/2}$. 
However, physics at $u\ne 0$
(in the ``bulk'') is not invariant under this transformation. 

The complete metric in the boosted system is given by the tensor 
transformation 
law 
\be{}
g'_{\alpha \beta}=\frac{\partial x^\mu}{\partial x'^\alpha}
\frac{\partial x^\nu}{\partial x'^\beta}\,g_{\mu \nu},
\ee
where $g_{\mu \nu}={\rm diag}(1,-\alpha^2,-1)$ is the Gauss-normal 
metric of Eq.~\rf{metric}.
Using Eq.~\rf{ltransinv}, the resulting boosted metric is  
\be{}\label{boostedmetric}
g'_{\mu\nu}=
\left(
\barr{ccr}
\gamma^{2}(1-\beta^{2}\alpha^2) 
    & \gamma^{2}\beta(1-\alpha^2) & 0 \\
 \gamma^{2}\beta(1-\alpha^2) 
& -\gamma^{2}(\alpha^2-\beta^{2}) & 0 \\
0 & 0 & -1 
\earr
\right).
\ee
Notice that only for $\alpha^2=1$ is the metric Lorentz invariant.  
Such is the case on our brane, but generally not the case on other 
hypersurfaces.  On other hypersurfaces, Eq.~\rf{metric} gives the
limiting velocity seen by local inhabitants
in the rest frame 
as $\alpha^{-1}(u_j) \equiv \alpha^{-1}_j$. However, this value is 
not invariant under Lorentz boosts defined on our brane.

At first glance the metric (\ref{boostedmetric}) seems to belong to the 
to the broad class of metrics~(\ref{stockum}),
which includes the G\"odel- and Tipler-van-Stockum (GTvS) spacetimes. 
After all, $g_{tx}\ne 0$ where $\alpha\ne 1$, i.e., off the brane.
Moreover, ${\rm sign} (g_{xx})$ is adjustable.
However, a significant difference from the GTvS metric is that,
in the case here of a boosted 
asymmetrically warped extra dimension, the variable $x$ is not periodic
(unless our universe has the topology of a flat torus).
It is thus required to construct an explicit return path to the spacetime
point of origin. 
We will now show that it is not possible to construct a return path that 
arrives 
sufficiently quickly to close the curve.

The first way we show the absence of a CTC is to inject the metric elements 
of~\rf{boostedmetric} into the CTC conditions~\rf{CTCconditions}.
A boost does not alter the determinant of the metric, 
so $\sqrt{g_5}=|\alpha|$.
With $g_{tx}=\gamma^2\beta(1-\alpha^2)$, and 
$g_{xx}=\gamma^2(\alpha^2-\beta^2)$,
one finds that the CTC condition in Eq.~\rf{CTCcondition1} 
is satisfied
only
if $|\alpha_1| < -|\alpha_2|$,
which is a contradiction.
So there is no CTC. 

Secondly, we show explicitly that the physics failure arises from 
unavailability of 
any return path to close the curve.
For our present 5D discussion, the path is that of Fig.~\ref{bulkpath} 
when one ignores the sixth dimension coordinate $v$.
The signal leaves 
our brane at the spacetime point $O=(t=0,x=0, u=0)$ and arrives 
at $u=u_1\ne 0$, 
and then propagates on the 
hypersurface at $u_1$ for a travel time $t$ with the
limiting velocity $(\alpha(u_1))^{-1}\equiv\alpha_1^{-1}$.
We will assume that $0 < \alpha_1 <1$, so that the travel speed 
in the bulk 
is superluminal relative to travel speed on our metric. 
At time $t$, 
the signal may 
reenter our brane.
In the limit $u_1\ll \alpha_1^{-1}t$,
which is always fulfilled for sufficiently large $t$, 
the reentry point on our brane is $B^\mu\approx (t,x=\alpha_1^{-1}t, u=0)$.
Since the distance to the reentry point $B^\mu$ is spacelike
(i.e. outside the brane's light-cone), 
it may be transformed to negative time by a boost on our brane.  
The boosted point $B'^{\mu}$ is obtained by using the 
transformation (\ref{ltrans}).  The point $B'^{\mu}$ has coordinates
\be{}\label{ltrans2}
x'=\gamma\,t\left(\alpha_1^{-1}-\beta\right),~~~~
t'=\gamma\,t\left(1-\beta \alpha_1^{-1}\right).
\ee
It is clear that for
\be{}
0<\alpha_1<\beta<1 \qquad [{\rm equivalent\ to\ }g_{xx}(u_1)<0]
\ee
an observer in the boosted frame on our brane sees the signal arrive  
in time with $t'<0$, i.e., before it was emitted.  
However,
this result alone does not imply any conflict with causality.
In particular, it does not necessarily imply that spacetime is blessed with 
CTCs. 
To close the timelike curve, 
one has to show that the time $t'$ during which 
the signal traveled backwards in time, 
is sufficiently large to allow a return from the
spacetime point $B'^{\mu}=(t',(x=\alpha_1^{-1}t)',0)$ on our brane to the 
spacetime point of origin, $O=O'=(0,0,0)$. 
The speed required to close the lightlike curve of the signal,
as seen by the boosted observer on the brane, is 
\be{}
\label{creq}
c'_{\rm req}=\frac{(x=\alpha_1^{-1}t)'}{|t'|}
=\frac{1-\beta \alpha_1}{\beta-\alpha_1}\,,
\ee
where the latter expression results from inputting Eq.~\rf{ltrans}.
It is easy to show that the condition 
$0<\alpha_1<\beta<1$ implies that $c'_{\rm req}$ itself 
is superluminal.
Thus there is no %
return path on our brane which leads %
to a CTC.
To generate a CTC 
the signal has to traverse another 
path (say, at constant $u_2$) which has a
limiting velocity satisfying $c'_{\rm bulk} \geq c'_{\rm req}$ in the 
$v$-frame.

Using the general expression for the metric in Eq.~\rf{boostedmetric}, 
the null line element for this hypersurface is, 
with $\alpha_2\equiv \alpha(u_2)$,
\be{} 
\label{boostedlinement}
0= ds'^2=\gamma^{2}\left\{
\left(1-\beta^{2} \alpha_2^2 \right)dt'^2
+ 2 \beta\left(1-\alpha_2^{2}\right)dx' dt'
-\left(\alpha_2^2-\beta^{2} \right)dx'^2
\right\}\,.
\ee
There results a quadratic equation for $c'_{\rm bulk}$.
From Table~\ref{table:velocities} we see that the only possible solution is
${\dot x}_-$, given by Eq.~\rf{velocities1} with $v$ set to zero and $-g_6$ 
replaced by $g_5=\alpha_2^2$. 
Thus we have
\be{}
c'_{\rm bulk-}\equiv {\dot x}'_- = \left( \frac{dx'}{dt'} \right)_-
= \frac{\gamma^2\beta(1 - \alpha_2^2)-\alpha_2}{\gamma^2(\beta^2-\alpha_2^2)}
=\frac{1+\beta\alpha_2}{\beta+\alpha_2}\,.
\label{cbulk}
\ee
It is relatively easy to see that this result satisfies the inequality chain 
$|c'_{\rm bulk-}|< |1/\beta| < |c'_{\rm req}|$.
(One way to see this inequality chain is to recognize that 
the RHS of~\rf{cbulk} has the form of the velocity addition formula in 
Special Relativity,
where the velocity sum is bounded by unity when the ``velocities'' 
$\beta$ and $\alpha_2$ 
are themselves bounded by unity.)
Thus, the return path cannot 
be superluminal and thus cannot
close the timelike curve,
and there is no CTC.

So we conclude that in an asymmetrically warped space 
with only one extra 
dimension (\ref{metric}), as well as its boosted equivalent
(\ref{boostedmetric}), no CTCs exist 
if space dimensions are non-periodic.
5-dimensional brane universes with one asymmetrically
warped extra dimension are causally stable.
(The exception is the topology of a flat
torus, which maps spacetime into the class of GTvS 
spacetimes of Eq.~(\ref{stockum}).)
We are thus led to consider next a spacetime with two asymmetrically
warped extra dimensions.  There we will find that CTCs do exist.
The lesson learned from the attempt to formulate a 5D metric having
CTCs will provide 
intuitive input into the construction of the 6D metric.

\section{CTCs with two warped extra dimensions}
\label{sec:6D}

We now proceed by constructing a 6D metric exhibiting CTCs, which is a 
natural generalization of the metric (\ref{metric}). 
Let  ``$u$'' and ``$v$'' label the two extra space dimensions.
We assume that these dimensions have warp factors $\alpha(u)$ and $\eta(v)$,
respectively. 

In our attempt to construct an asymmetrically warped 
metric exhibiting CTCs in 5D, 
we found that the 
metric in Eq.~\rf{metric} allowed a quantum to travel superluminally
into the bulk. Being outside our light-cone, the worldline of this
quantum could be boosted to negative time by a Lorentz transformation 
on the brane. However, we showed that a superluminal return path
to the brane was required to close the worldline and there was no such path.
This failure can be traced to the fact that the Lorentz transformation
was just a coordinate change, and so provided a change of view, but no new 
physics. What is needed is a nonzero $g_{tx}$ that cannot be
removed by a linear transformation among brane coordinates.
Introducing the 6th dimension provides a solution, first because it allows
a superluminal return path along the additional 6th dimension, and second
because it allows 
$g_{tx}(u,v)$ to be ``hard-wired'' into the metric so that it is
not removable by a linear coordinate transformation on the brane.
 (Recall that we learned in Section~\rf{sec:extendingST}, 
via Eqs.~\rf{CTCconditions} and the discussion just below these,
that a nonzero $g_{tx}(u,v)$ is a necessary ingredient for the 
existence of CTCs.)

A natural 6D generalization of (\ref{metric}) can be realized by assuming
that the metric for the $u$- and $v$ dimensions exhibits the 
simple form in~(\ref{metric}), but in different Lorentz frames. 
This assumption seems natural for any spacetime with two or more extra 
dimensions, 
since there is no preferred Lorentz frame for the bulk, from the viewpoint 
of the brane. In analogy to (\ref{metric})
it could be realized
by assuming two AdS-Schwarzschild or AdS-Reissner-Nordstr\"om black holes
being located in the $u$ and $v$ dimension and moving with a relative 
velocity.
This choice also ensures superluminal travel to as well as from the brane, 
as well as a Minkowskian metric on the brane.
To construct this 6-dimensional metric explicitly,
let us denote by $\beta_{uv}$ 
the ``relative velocity'' between the two Lorentz frames 
in which the $u$ and $v$ dimensions assume the
simple form~(\ref{metric}), respectively.
We incorporate the ``$u$-frame'' slice at $v=0$ by
retaining the warp factor $\alpha(u)$
on the brane coordinate $dx$, and we incorporate the ``v-frame'' slice
at $u=0$ by writing the boosted metric in Eq.~\rf{boostedlinement} 
with the warp $\alpha(u)$ now replaced by $\eta(v)$. The resulting
full 6-dimensional metric then has the form
\be{}
\label{6dmetric}
ds^2=\gamma_{uv}^2 \left\{\,
[1-\beta_{uv}^2 \eta^2(v)] \,dt^2 + 2 \beta_{uv} \alpha(u) 
[1-\eta^2(v)] \,dx 
dt
-\alpha^2(u) [\eta^2(v) - \beta^2_{uv}] \,dx^2 \,
\right\} - du^2 - dv^2.
\ee
One easily finds that $-Det\equiv-g_6=\alpha^2(u)\,\eta^2(v)$.
That this determinant is independent of $\beta_{uv}$ is consistent 
with the interpretation of $\beta_{uv}$ as a kind of boost parameter.
Of special importance for the existence of 
the CTC is the off-diagonal metric element $g_{tx}$,
which is nonzero for $\eta(v)\ne 1$ (i.e., off the brane), 
and the metric element $g_{xx}$ which is of indeterminate sign.
As a consistency check on the metric, we note that for $u=v=0$, 
i.e., on the brane, 
Eq.~\rf{6dmetric} reduces to 4-dimensional Minkowski spacetime.

\begin{figure}
\centering
\includegraphics[clip,scale=0.50]{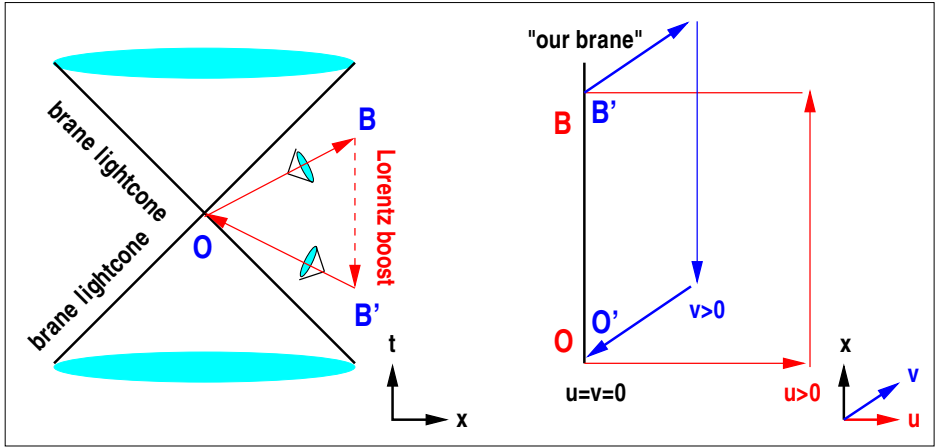}
\caption{
Closed timelike curve in an asymmetrically warped universe:
(i) A signal takes a spacelike shortcut via a path of constant
$u>0$ with $v=0$ from point $O$ to point $B$.
(ii) A Lorentz boost transforms $B$ into $B'$ with negative time coordinate. 
(iii) A return shortcut at constant 
$v>0$ with $u=0$ closes the timelike curve.
The return path to the brane and the boost between steps (i) and (iii) is 
pedagogic rather than necessary, and an intermediate path in the bulk 
connecting the $u>0$ and $v>0$ hyperslices is possible. 
(Note: The tipped light-cones in the figure are symbolic rather
than quantitative.) 
}
 \label{bulkpath}
\end{figure}

We now present two arguments, parallel to those given in Sec.~\rf{sec:5D} for 
the 5D case,
but leading to the opposite conclusion, namely that the metric~\rf{6dmetric} 
does support CTCs.

The first argument establishing the existence of 
a CTC is to show that the metric elements in 
\rf{6dmetric} can be chosen to satisfy the two CTC conditions of 
Eq.~\rf{CTCconditions}.
Inputting the metric elements into~\rf{CTCconditions},
one finds, after some algebra, 
that the conditions reduce to 
\beq{6dCTCa}
\frac{\alpha_2\,(\beta_{uv}+\eta_2)}{1+\beta_{uv}\eta_2} 
< \frac{\alpha_1\,(\beta_{uv}-\eta_1)}{1-\beta_{uv}\eta_1}\,,
\ee
and
\beq{6dCTCb}
\eta_1<\beta_{uv}.
\ee
The new feature here, as opposed to the 5D metric, is the freedom to choose 
$\alpha_1$ and $\alpha_2$ to ensure that the CTC conditions are satisfied.
We see that any pair $(\alpha_1,\alpha_2)$ will do, as long as they satisfy
\beq{alphacondition}
\frac{\alpha_2}{\alpha_1} < 
\left( \frac{\beta_{uv}-\eta_1}{1-\beta_{uv}\eta_1}\right)\,
   \left(\frac{1+\beta_{uv}\eta_2}{\beta_{uv}+\eta_2}\right)\,.
\ee
This inequality can always be satisfied by an arbitrarily small choice for 
$\alpha_2$.

One simple and successful choice is to set $\alpha_1=1$ and $\eta_2=1$, 
i.e., to take the outgoing path on the $u=0$ hyperslice
and the return path on 
the $v=0$ hyperslice
(and the steps (i) and (iii) in 
Fig.~\ref{bulkpath} are interchanged).
With these choices, (\ref{alphacondition}) reduces to
$\alpha_2<(\beta-\eta_1)/(1-\beta \eta_1)$. This is guaranteed to be 
satisfiable by (\ref{6dCTCb}).
The choices $u_1=0$ and $v_2=0$ will lead to an
explicit CTC.
With $u=0$, Eq.~\rf{6dmetric} reduces to~\rf{boostedlinement} with $\eta^2(v)$ 
replacing $\alpha^2(u)$:
\beq{metricu0}
ds^2|_{u=0}=\gamma_{uv}^2 \left\{\,
[1-\beta_{uv}^2 \eta^2(v)] dt^2 + 2 \beta_{uv} [1-\eta^2(v)] dx dt
 -[\eta^2(v) - \beta^2_{uv}] dx^2\,
\right\} - dv^2.
\ee
Thus we see explicitly that choosing $\eta_1<\beta_{uv}$ on the $u=0$
 hyperslice sets $g_{xx}<0$, 
so that our outgoing path necessarily accumulates negative time
(original frame in Table~\ref{table:time-running}).
On the return path, we set $v=0$.  
Then the 6D metric of Eq.~\rf{6dmetric} reduces to \rf{metric}, repeated here:
\beq{metricv0}
ds^2|_{v=0}=dt^2-\alpha^2(u) dx^2-du^2\,;
\ee
It is clear that this return path can be made arbitrarily brief by choosing 
$\alpha_2$ 
arbitrarily small.
The CTC is revealed.

We note that when the metric~\rf{6dmetric} is transformed into the $v$-frame
by a Lorentz transformation on the brane
with $\beta=-\beta_{uv}$,
then the metric along the $v$-dimension assumes the simple form 
of~\rf{metricv0} 
(with obvious replacements) and the 
metric along the $u$-dimension becomes non-diagonal
($v$-frame in Table~\ref{table:time-running}).

\begin{table}
\centering
\begin{tabular}{|c|c|c|} \hline
                &  original $u$-frame       &  (un)-boosted $v$-frame \\
\hline
\hline
 $u\neq 0$      &  forward in time      &  backwards in time  \\ \hline
 $v\neq 0$      &  backwards in time    &  forward in time     \\ \hline
CTC &&\\
            &   \includegraphics[clip,scale=0.30]{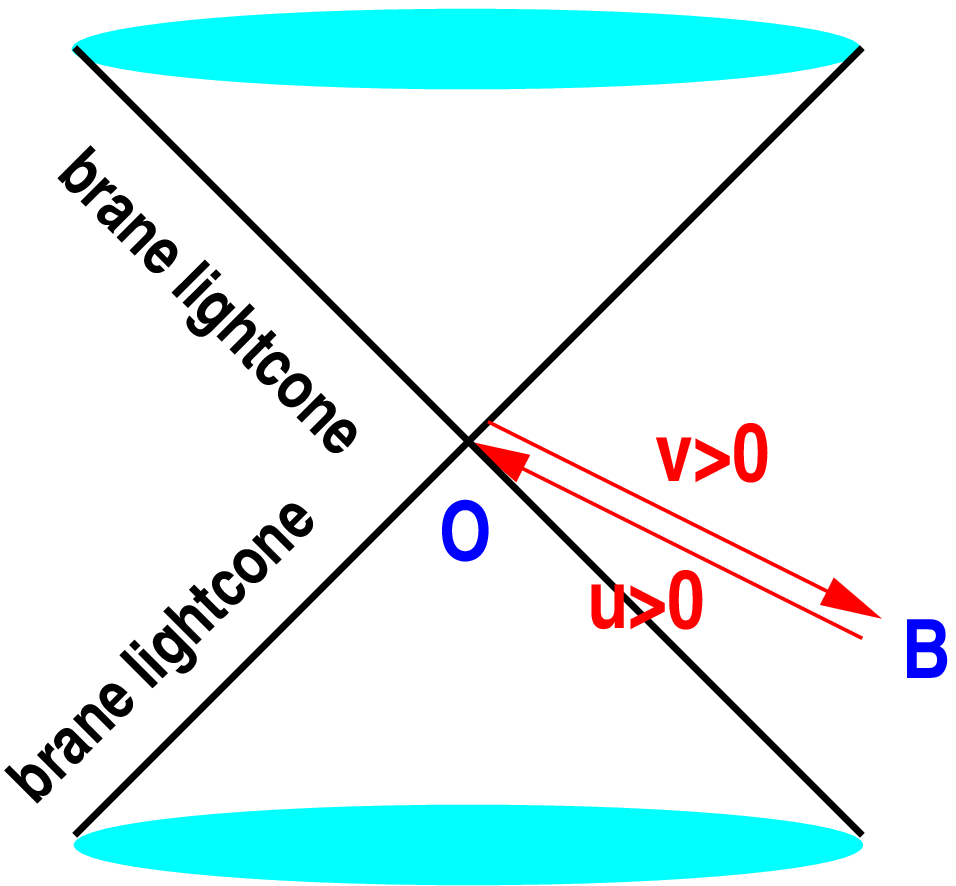}
 &   \includegraphics[clip,scale=0.30]{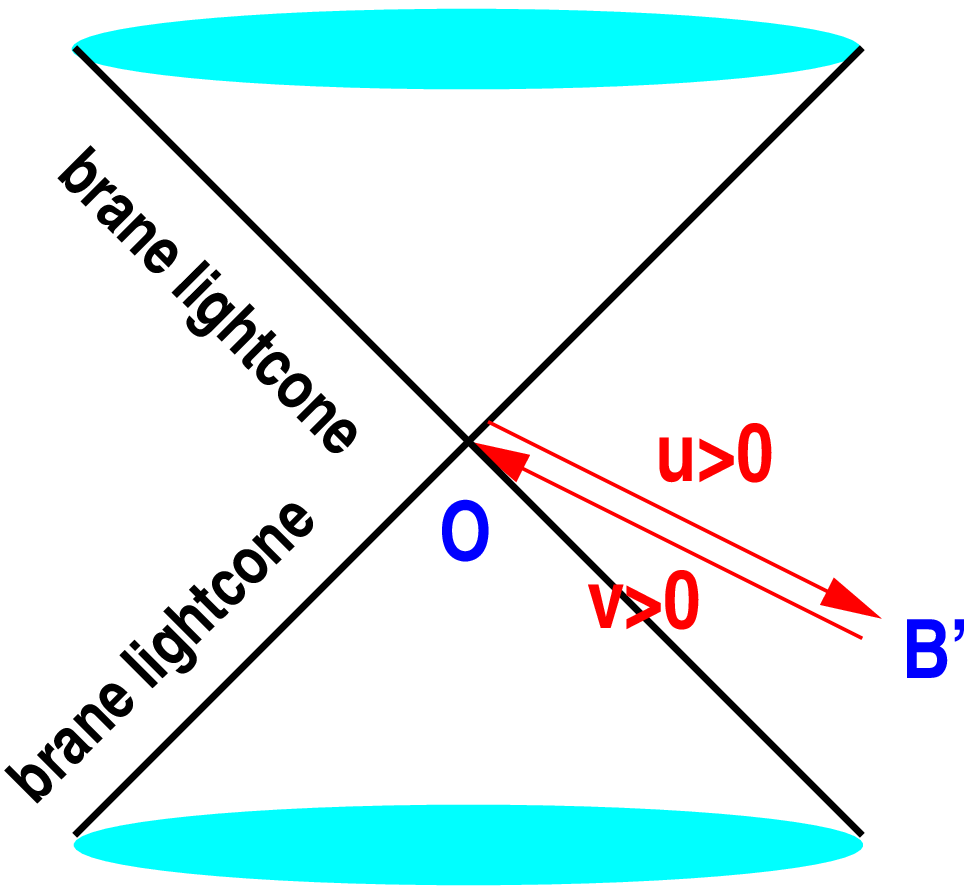}                    
\\ \hline
\end{tabular}
\caption{ 
\label{table:time-running}
Running of time in the original frame, where the metric assumes the form
(\ref{6dmetric}), and in the frame where the metric is boosted into the 
$v$-frame with
$\beta=-\beta_{uv}$.
} 
\end{table}

To summarize this section, we have identified a CTC 
beginning and ending on our brane and superluminally transiting
two paths parallel to our brane but in the asymmetrically warped 
$u$- and $v$-dimensions. 
The physics that enables the CTC is the breaking of global Lorentz 
invariance away from the brane.

\section{Stress-energy tensor and energy conditions}
\label{sec:Tmunu}
As a check on the consistency of the picture, 
we should diagnose the stress-energy tensor 
which sources the extra-dimensional metric, for any pathologies.
In particular, we will be interested in the resulting matter distributions on 
and off the brane.
Thus, our task is to calculate the Einstein tensor
\beq{Eisntein}
G_{\mu \nu}=R_{\mu \nu} -\frac{1}{2} g_{\mu \nu} R,
\ee
from the spacetime metric of Eq.~(\ref{6dmetric}),
and then to obtain the stress-energy tensor $T_{\mu\nu}$ via the Einstein 
equation
\beq{Tmunu}
T_{\mu \nu}=\frac{1}{8\,\pi\,{\rm G_N}}\,G_{\mu \nu}\,.
\ee
Consequently, we proceed to evaluate 
$T_{\mu\nu}=(8\pi\,G_{\rm N})^{-1}\,G_{\mu\nu}$
with no preconceptions as to its form.
We note that in general, 
$T_{\mu\nu}$ contains contributions from matter, fields, and cosmological 
constant 
on and off the brane, and from brane tension on the brane.

\begin{figure}
\centering
\includegraphics[clip,scale=1.]{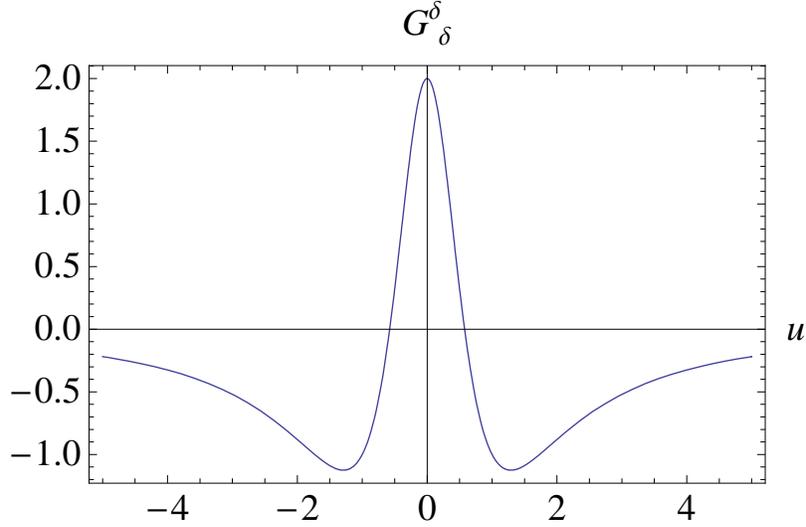}
\caption{
Nonzero elements of the Einstein tensor $G^\mu_{\ \nu}$ (in arbitrary units):
$G^\delta_{\ \delta}\equiv G^0_{\ 0}=G^y_{\ y}=G^z_{\ z}=G^v_{\ v}$,
on the $v=0$ slice, as a function of $u$.
Assumed are
warp factors $\alpha(u)=1/(u^2 + c^2)$ and $\eta(v)=1/(v^4 + c^2)$, 
with $c=1$. We find that the weak and dominant energy
conditions are violated in the bulk, while all energy conditions with
the exception of the SEC are satisfied on the brane.}
\label{encond}
\end{figure}

%
%
%

Instead of complicated analytic expressions for $T_{\mu\nu}$, 
we present some visual output~\cite{mathpackage} 
of the Einstein tensor versus $u$, on the $v=0$ slice.  
We do so for warp factors $\alpha(u)$ and $\eta(v)$ 
chosen to satisfy energy conditions discussed below.
An analogous figure is the Einstein tensor versus $v$, on the $u=0$ slice.
However, this Einstein tensor has off-diagonal elements,
which increases the number of figures.
Furthermore, it offers us no additional enlightenment,
so we do not show this Einstein tensor.

%
%

There is considerable theoretical prejudice that stable Einstein 
tensors should 
satisfy certain ``energy conditions'' relating energy density 
$\rho$ and directional pressures $p^j$.
The null, weak, strong and dominant energy conditions state that 
\bea{energy}
{\rm NEC}&:& \rho + p^j \geq 0, ~~~\forall j\,. \\
{\rm WEC}&:& \rho \geq 0\,; ~~~{\rm and}~~~  \rho + p^j \geq 0, 
~~~\forall j\,. \\
{\rm SEC}&:& \rho + p^j \geq 0, ~~~ \forall j; ~~~{\rm and}~~~  
\rho + \sum_j p^j \geq 0\,. \\
{\rm DEC}&:& \rho \geq 0\,; ~~~{\rm and} ~~~  p^j \in [\rho,-\rho], 
~~~ \forall j\,.
\eea

For the purpose of definiteness in the identification of $\rho$ and $p^j$, 
we assume the anisotropic fluid relations
\beq{ideal fluid}
T^{\mu\nu}=-p\,g^{\mu\nu}+(\rho+p)\,U^\mu U^\nu\,,
\ee
with $u^\mu=(1,{\vec 0})$ being the net four velocity of the fluid.
The usual approach is to work with one raised and one lowered index to express 
the stress-energy in terms of metric $g_{\mu\nu}$ rather than inverse metric 
$g^{\mu\nu}$.  One has 
\beq{updown}
T^\mu_{\ \nu}=-p\,\delta^\mu_\nu+(\rho+p)\,g_{\nu\alpha}\,U^\mu U^\alpha\,.
\ee
Then, with a diagonal metric with $g_{tt}=1$ (Gaussian-normal coordinates),
one obtains for the nonzero elements of $T^\mu_{\ \mu}$,
\beq{rhoandp}
\rho=T^0_{\ 0} \quad{\rm and} \quad p^j=-T^j_{\ j}\,.
\ee
These are the relations appropriate for the $v=0$ slice of our metric,
since one sees in Eq.~\rf{metricv0} that the $v=0$ metric is manifestly diagonal with $g_{tt}=1$.

It is not difficult to find a functional form for the
warp factors $\alpha$ and $\eta$ which conserves some of 
the energy conditions, at least on the brane. 
One such example is given by 
$\alpha(u)=1/(u^2 + c^2)$ and $\eta(v)=1/(v^4 + c^2)$. For this case 
the 
elements of the Einstein tensor on the $v=0$~slice
are shown as a function of $u$ in Fig.~\ref{encond}. 
The null, weak and dominant energy conditions are
conserved on the brane, while the strong energy condition is 
violated both on the brane and in the bulk.  

The negative energy density that afflicts many wormhole and CTC 
solutions in four dimensions
is avoided on the brane in the example for an extra-dimensional 
CTC presented here.
However, $\rho$ becomes negative as one moves away from the brane into the 
bulk, so that the WEC and DEC are violated off the brane, while the NEC
remains satisfied.
We have successfully constructed a metric exhibiting CTCs 
in an extra-dimensional spacetime 
by "moving" the negative energy density from the brane to the bulk.
One might even speculate that the negative energy density in the bulk 
is related to the compactification of the extra dimensions, or 
possibly to the repulsion of Standard matter from the bulk.

One also sees in Fig.~\rf{encond} that $G^y_{\ y}=G^z_{\ z}=G^v_{\ v}$  
are equal to $G^0_{\ 0}$ on the $v=0$ slice.
This equality amounts to a dark energy or cosmological constant equation of state for the 
$y$-, $t$-, and $v$-directed pressures, namely,  $w^j\equiv p^j/\rho=-1$.
There may be some intriguing physics underlying this result.

\section{Discussion and Conclusion}
\label{sec:conclusion}
We have derived the general conditions on metric elements which allow 
spacetimes to 
contain closed timelike curves (CTCs).
Then, we have demonstrated the existence
of CTCs for a rather generic spacetime with
two asymmetrically warped extra dimensions.
In addition, we have found particular warp factors for the metric which yield
 positive energy
density on the brane. However, negative energy density is not completely 
banished,
as it does appear in the bulk. Since one cannot observe the bulk energy 
density,
we may at least say that negative energy density is banished from sight.
It is also possible that an anthropic argument applies here:
Life may evolve only where energy density is positive.
Then lifeless bulk regions of negative energy density can communicate their 
existence 
to living beings only via geometry, perhaps mediated by 
the exchange of gravitons or
appropriately named, ``sterile'' neutrinos.

It should be stressed that realistic graviton
or bulk fermion signals, rather than following restricted bulk trajectories 
with constant $u$ or $v$ as constructed here, 
will instead propagate on the path of least action to minimize
the travel time. Since the effectively
superluminal velocities in our constructed example 
produced a CTC, we expect that a truly geodesic signal
will also generate a CTC.
In this case the causal structure of extra dimensions may be
studied with sterile neutrino beams by utilizing 
resonant conversion of active neutrinos via matter effects into sterile
neutrinos and back.
We note that the model presented herein is complete in that the geodesic 
equations of motion are derivable from the metric in Eq.~\rf{6dmetric}.
We have not investigated the geodesic equations in this work.

A thorough discussion of whether
CTCs in the observable universe are hidden behind  chronology horizons 
where the stress-energy tensor
diverges (one may consult the discussion in \cite{kao,diaz}),
is beyond the scope of this work.
We have confined ourselves to the pragmatic attitude that even if
chronology were protected by some mechanism operative near the
chronology horizon, it remains a highly
rewarding effort to study the physics near
this horizon.  
The CTC we have constructed is particularly interesting
in this respect, since 
it could be available to gauge-singlet  
particles which have previously been 
hypothesized to propagate in the extra-dimensional bulk
\footnote{A realistic description of such particles taking shortcuts in
extra dimensions would require a quantum field theoretic treatment 
similar to the one preformed in \cite{Burgess:2002tb}.}.

\section*{Acknowledgments}
We thank 
the referee, whose constructive criticism improved the paper considerably.
We also thank
Luis Anchordoqui,
Roman Buniy,
Cliff Burgess,
Daniel Chung,
Joshua Erlich,
John Friedman,
Alan Guth,
Daniel Hobbs,
Hideki Ishihara,
Nick Kaiser,
Thomas W. Kephart,
Jeff Kuhn,
John G. Learned,
Ralf Lehnert,
Kirill Melnikov,
Octavian Micu,
Alexander Mitov,
Graham Shore,
Xerxes Tata,
Matthew Weippert,
and
T.-C\@. Yuan
for many critical comments and useful discussions. 
We thank the Tokkuri Tei establishment and especially Kazu Mitake
for providing a stimulating environment in which to further this paper.
This work was supported in part by US DOE under the grants DE-FG02-04ER41291
and DE-FG05-85ER40226.
TJW thanks the Alexander von Humboldt Foundation for support with a Senior
Research Award.
HP and TJW thank the Kavli Institute for Theoretical Physics China for
kind support and hospitality during the last stages of this work.

\end{document}